\documentclass[twocolumn,aps,prb]{revtex4}
\usepackage{amsfonts}
\usepackage{amsmath}
\usepackage{amssymb}
\usepackage{graphicx}
\begin{document}

\title{Statistics of spinons in the spin-liquid phase of Cs$_2$CuCl$_4$}

\author{Chung-Hou Chung}
\author{Klaus Voelker}
\author{Yong Baek Kim}
\affiliation{Department of Physics, University of Toronto, Toronto, Ontario, 
Canada M5S 1A7}
\date{\today}

\begin{abstract}
Motivated by a recent experiment on 
Cs$_2$CuCl$_4$,\cite{coldea}
we study the spin dynamics of the spin-liquid phase of the spin-1/2
frustrated Heisenberg antiferromagnet on the anisotropic triangular 
lattice. There have been two different proposals\cite{chung,zhou} for 
the spin-liquid phase of Cs$_2$CuCl$_4$. These spin-liquid states 
support different statistics of spinons;
the bosonic Sp($N$) large-$N$ mean field theory\cite{chung,subir} 
predicts bosonic spinons, while the SU(2) 
slave-boson mean field theory\cite{zhou,wen02} leads to fermionic
spinons. We compute the dynamical spin structure factor for both types of 
spin-liquid state at zero and finite temperatures. 
While at zero temperature both theories agree with experiment on a
qualitative level, they show substantial differences in the temperature
dependence of the dynamical spin structure factor.
\end{abstract}

\maketitle

The existence of a spin-liquid state in frustrated quantum magnets
has been one of the central issues in the field of strongly correlated
systems.\cite{anderson73,sondhi01,balents01,nayak01,capriotti,z2}
Recent experimental studies of various frustrated magnetic
compounds provided excellent opportunities to investigate the competition
between geometric frustration and quantum fluctuations, and its role
in the occurrence of quantum spin-liquid states.\cite{ramirez}
Much interest in two-dimensional quantum spin-liquid
states has evolved since Anderson proposed that the doped spin-liquid state may hold
the key to the puzzles of high temperature superconductivity 
in Cuprates.\cite{anderson87}
More recently the search for spin-liquid states has been extended
to geometrically frustrated quantum
magnets.\cite{sondhi01,balents01,nayak01,capriotti}

The hallmark of the spin-liquid
state is the existence of fractionalized excitations. 
In this regard, the two-dimensional frustrated 
Heisenberg antiferromagnet Cs$_2$CuCl$_4$ provides a useful
realization of a two-dimensional spin-liquid state:
A recent neutron scattering experiment \cite{coldea} showed the remarkable
result that the dynamical spin structure factor 
${\cal S}({\bf q},\omega)$ does not exhibit well-defined peaks corresponding 
to spin-1 magnons. Instead, there exists a continuum of
excitations which has been interpreted as the indication of
pairs of deconfined spin-1/2 spinons in the underlying spin-liquid state.

The minimal Hamiltonian describing Cs$_2$CuCl$_4$ is argued to be 
the spin-1/2 Heisenberg antiferromagnet on an anisotropic
triangular lattice:\cite{coldea}
\begin{eqnarray}
H &=& 
J_1\sum_{<ij>} {\bf S}_{i}\cdot {\bf S}_{j} + 
J_2\sum_{<<ij>>} {\bf S}_{i}\cdot {\bf S}_{j}, 
\label{H}
\end{eqnarray}
where ${\bf S}_{i}$ is the $S = 1/2$ spin operator on
site $i$, and $J_1$, $J_2$ are the exchange couplings 
along two different types of bonds, as indicated in Fig.\ 1.
There have been at least two different proposals for the spin-liquid
state of this model in connection to the experiment on Cs$_2$CuCl$_4$.
These proposals are based on two different mean field approaches:
the bosonic Sp($N$) large-$N$ mean field theory\cite{chung,subir} and 
the SU(2) slave-boson mean field theory.\cite{zhou,wen02,su2}
One of the distinguishing properties of the two resulting spin-liquid states
is the statistics of the fractionalized excitations, {\it i.e.}, spinons;
the bosonic Sp($N$) large-$N$ mean field theory supports bosonic spinons
while the SU(2) slave-boson mean field theory leads to fermionic spinons.

\begin{figure}[hb]
\includegraphics[width=4.0cm,angle=0]{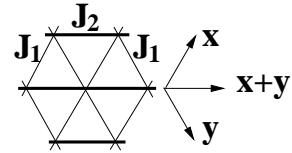}
\caption{The antiferromagnetic exchange interactions in Cs$_2$CuCl$_4$ 
with $J_1 = 0.125$meV and $J_2 = 0.375$meV.\cite{coldea} }
\label{lattice}
\end{figure}

The statistics of the spinons may be a useful characteristic of 
the underlying spin-liquid states. In particular, an important question
is whether spin-liquid ground states with the same symmetry, but 
different spinon statistics, are necessarily distinct;
this question has been only partially answered.\cite{demler}
In this paper, we would like to achieve a more moderate goal; while the 
above question about the ground state cannot be answered by investigating 
mean field theories, one may still be able to identify 
which mean-field state provides a more faithful approximation of the true
spin-liquid state of the material. If the mean-field state is a good 
representation, residual quantum fluctuations will be small, and are not
expected to influence the system's responses in a qualitative way. 
We aim to achieve this goal by
comparing the spin excitation spectra of the different
proposals with the experimentally measured spin structure factor. 
We compute the dynamical spin structure factor for the spin-liquid
states in both the bosonic Sp($N$) large-$N$ and the SU(2) slave-boson
mean field theories. 
We find that, at temperatures well below the temperature scale set by
the exchange couplings, the temperature evolution of the dynamical 
spin structure factor depends significantly on the spinon statistics:
While the fermionic spinons are fairly insensitive to temperature,
the bosonic spinon spectrum shows significant changes in this temperature
range, as described below.

A useful bosonic representation of the Hamiltonian in Eq.\ \ref{H}
can be obtained by generalizing the physical spin 
SU(2) $\cong$ Sp(1) symmetry to Sp($N$). The generalized
spin operators can be expressed in terms of boson operators
$b_{i \alpha}^{\dagger}, b_{i \alpha}$ on every site $i$, where
$\alpha = 1 \ldots 2N$ is a Sp($N$) index.\cite{subir} 
The constraint $b^{\dagger\alpha}b_{\alpha} = n_b$ is imposed 
on each site to fix the number of bosons ($n_b = 2 S$ for $N=1$).
In the mean-field theory, the Sp($N$) Hamiltonian is solved 
in the $N\to\infty$ limit with $\kappa = n_b/N$ fixed.\cite{subir}
The mean-field phase diagram 
as a function of $J_2/(J_1+J_2)$ and $1/\kappa$ contains 
both magnetic long-range-order (LRO) and short-range-order (SRO) 
phases  (see Fig.\ 4 of Ref.\ \onlinecite{chung}) at zero temperature.
For bare parameter values relevant to 
the material ($\kappa = 1$ and $J_2/J_1 = 3$), the large-$N$ phase diagram 
predicts a spin-ordered ground state with an incommensurate
wavevector in the $J_2$-direction.\cite{chung}
However, finite-$N$ corrections will
move the phase boundaries, so that the physical spin-1/2 limit could in fact be
described by the incommensurate SRO (spin-liquid) phase 
with deconfined spin-1/2 spinons.\cite{chung} 
The material Cs$_2$CuCl$_4$ actually exhibits long-range order at 
temperatures below 0.62K, which can be suppressed by an applied 
magnetic field. Since this ordering is due to the small interlayer
coupling ($J_z < 10^{-2}J_2$),\cite{coldea} we expect the excitation
spectrum to be indistinguishable from the disordered state at energies
above this (small) scale. In our strictly two-dimensional model, no
ordered state can occur at any finite temperature.
Here, we consider
the point $\kappa = 0.64$ and $J_2/J_1 = 3$ in the large-$N$ 
phase diagram
(see Fig.\ 4 of Ref.\ \onlinecite{chung}),
which lies in the SRO 
phase close to the LRO phase boundary, as the possible spin liquid 
state relevant to Cs$_2$CuCl$_4$,
which is experimentally known to be close to a quantum phase transition.
The spinon dispersion in this case 
has a small gap at ${\bf q} = (0.26\pi, 0.26\pi)$ of 
approximately $0.03J_2$, which is much smaller than the experimental resolution of 
about $0.5 J_2$.\cite{coldea}


On the other hand, spin-liquid phases can also be obtained from a fermionic
representation via the SU(2) slave-boson approach.\cite{zhou,wen02,su2}
This approach utilizes a hidden SU(2) gauge symmetry in the fermionic representation of the 
Heisenberg model.\cite{su2} 
Introducing two SU(2) doublets,
$\psi^T_{i1}=(f_{i\uparrow},f_{i\downarrow}^{\dagger})$ and
$\psi^T_{i2}=(f_{i\downarrow},-f_{i\uparrow}^{\dagger})$,
to rewrite the destruction operators of spin-up and down
states,\cite{su2}
the mean-field Hamiltonian can be expressed in terms of
the $2\times 2$ Hermitian matrices $u_{ij}$ as
\[
H_{\mathrm{mf}}=-\sum_{ij} (\psi^{\dagger}_{i \alpha}
u_{ij} \psi_{j \alpha} + h.c.) +
\sum_i a^l_0 \psi^{\dagger}_{i \alpha}
\tau^l \psi_{i \alpha}.
\]
Here, $\tau^{l}$ ($l=1,2,3$) are the Pauli matrices, and the Lagrange multiplier
$a_{0}^{l}$ is used to enforce the constraint
$\langle \psi_{i \alpha}^{\dagger}\tau^{l}\psi_{i \alpha}
\rangle = 0$.\cite{zhou,wen02}
It has been proposed that the spin-liquid
phase with the following mean-field ansatz is the most likely candidate
for the spin-liquid phase of Cs$_2$CuCl$_4$:\cite{zhou}
\begin{eqnarray}
u_{i,i+\hat{x}} &=& \chi \tau^2,\hspace{0.2cm} u_{i,i+\hat{y}} = -\chi\tau^2,
\hspace{0.2cm}u_{i,i+\hat{x}+\hat{y}} = \lambda \tau^3, \nonumber \\
a_{0}^{1,2} &=& 0,\hspace{0.4cm} a_{0}^{3} = a_3.
\end{eqnarray}
Here the parameters $\chi, \lambda$ and $a_3$ have to be determined
self-consistently. According to the classification scheme of Ref.\ \onlinecite{zhou},
this is one of the possible U(1) spin-liquid phases on the anisotropic
triangular lattice.
The excitation spectrum has gapless points\cite{zhou}
at $(0, 0)$, $(\pi, \pi)$, and $(q,q)$, where $q = \pm \pi/2 \pm \epsilon$,
with $\epsilon = 0.04\pi$.
Although the mean-field solution is expected to be modified
by gauge field fluctuations, it may still provide a
qualitatively correct description of the spin excitation spectrum.

\begin{figure}[h]
\includegraphics[width=8.5cm]{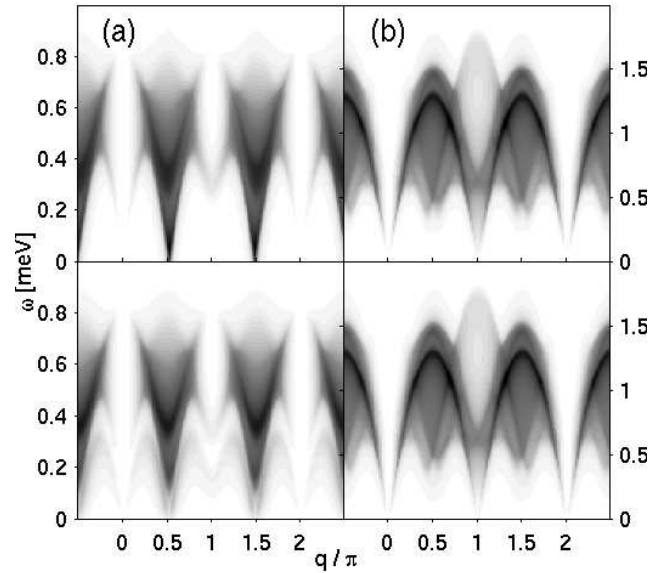}
\caption{Intensity plots of $\chi^{\prime\prime}({\bf q},\omega)$, with ${\bf q}$
oriented along the direction of $J_2$,
computed in (a) the Sp($N$) and (b) the SU(2) approach.
Darker areas indicate higher scattering intensity. 
Top panels represent the results at $T = 0$ and bottom panels indicate 
the results at $T = 0.15 J_2$. 
}
\label{denXSpN}
\end{figure}


The spin-liquid states obtained from the two approaches
differ in important aspects: one of them is a $Z_2$ spin-liquid
phase with gapped bosonic spinons (although the gap is small), and
the other is a U(1) spin-liquid state with gapless fermionic spinons.
It is therefore interesting to ask which spin liquid state is more
likely to describe Cs$_2$CuCl$_4$. We therefore calculate in both cases
the dynamical spin structure factor defined as
\(
  {\cal S}({\bf q},\omega) = -\frac{1}{\pi} (1 + n(\omega))
  \chi^{\prime\prime}({\bf q},\omega),
\) 
where
$n(\omega) = 1/(e^{\omega/T} - 1)$ is the Bose thermal factor,
and
$\chi^{\prime\prime}({\bf q},\omega)$ is
the imaginary part of the dynamical spin susceptibility:
\begin{equation}
\chi^{\prime\prime}({\bf q},\omega)
= 3 \mathrm{Im} \left ( -\hspace{-0.05cm}\int_{-\infty}^{\infty} \hspace{-0.3cm}dt
e^{i \omega t}\theta (t) \langle
[S^{z}({\bf q},t), S^{z}(-{\bf q},0)] \rangle \right ),
\end{equation}
where spin rotation invariance has been used to write $\chi = 3\chi_{zz}$.

In the Sp($N$) approach $\chi^{\prime\prime}({\bf q},\omega)$ 
is obtained as
\begin{eqnarray}
&&\hspace{-0.7cm}
\chi^{\prime\prime}_{{\rm Sp}(N)}
=  3 \sum_k \left [ (\frac{1}{4} + A({\bf q}))
(n(E_{\bf k}) - n(E_{{\bf k}}^{\prime}))
\delta (\omega + E_{\bf k} - E^{\prime}_{\bf k}) \right. \nonumber \\
&& + (\frac{1}{4} + A({\bf q}))
(n(E_{{\bf k}}^{\prime}) - n(E_{\bf k}))
\delta (\omega + E^{\prime}_{\bf k} - E_{\bf k}) \nonumber \\
&& + (\frac{1}{4} - A({\bf q}))
(1 + n(E_{\bf k}) + n(E_{{\bf k}}^{\prime}))
\delta (\omega + E_{\bf k} + E^{\prime}_{\bf k}) \nonumber \\
&& - \left. (\frac{1}{4} - A({\bf q}))
(1+n(E_{{\bf k}})+n(E_{\bf k}^{\prime}))
\delta (\omega - E_{\bf k} - E^{\prime}_{\bf k})] \right ],
\nonumber \\
\end{eqnarray}
where $A({\bf q}) = \frac{\bar{\lambda}^{2}
-\Delta_{\bf k}\Delta_{\bf k}^{\prime}}{4E_{\bf k} E^{\prime}_{\bf k}}$,
$E^{\prime}_{\bf k} = E_{{\bf k}+{\bf q}}$, and
$\Delta^{\prime}_{\bf k} = \Delta_{{\bf k}+{\bf q}}$.
Here $n(E_{\bf k})$ is the Bose thermal factor,
$E_{\bf k} = \sqrt{\bar{\lambda}^2 - \Delta^{2}_{\bf k}}$ is the bosonic
spinon dispersion\cite{chung}, and
$\Delta_{\bf k} = J_1 Q_1 (\sin k_x + \sin k_y) + J_2 Q_2 \sin(k_x + k_y)$.
$Q_1$ and $Q_2$ are mean-field bond variables for 
nearest-neighbor and next-nearest-neighbor bonds,
respectively, and $\bar \lambda$ is a Lagrange multiplier enforcing the constraint
on the number of bosons.
The temperature dependent values of $Q_1$, $Q_2$, and $\bar\lambda$ are determined
self-consistently.

In the SU(2) slave-boson approach, $\chi^{\prime\prime}$ is given by
\begin{eqnarray}
&& \hspace{-0.7cm}
\chi^{\prime\prime}_{SU(2)}
 = 3 \sum_k \left [ (\frac{1}{4} + B({\bf q}))
(f({\cal E}_{\bf k}) - f({\cal E}_{\bf k}^{\prime}))
\delta (\omega + {\cal E}_{\bf k} - {\cal E}^{\prime}_{\bf k}) \right. \nonumber \\
&& + (\frac{1}{4} + B({\bf q}))
(f({\cal E}_{\bf k}^{\prime}) - f({\cal E}_{\bf k}))
\delta (\omega + {\cal E}^{\prime}_{\bf k} - {\cal E}_{\bf k}) \nonumber \\
&& - (\frac{1}{4} - B({\bf q}))
(1 - f({\cal E}_{\bf k}) - f({\cal E}_{\bf k}^{\prime}))
\delta (\omega + {\cal E}_{\bf k} + {\cal E}^{\prime}_{\bf k}) \nonumber \\
&& + \left. (\frac{1}{4} - B({\bf q}))
(1 - f({\cal E}_{\bf k}) - f({\cal E}^{\prime}_{\bf k}))
\delta (\omega - {\cal E}_{\bf k} - {\cal E}^{\prime}_{\bf k}) \right ],
\nonumber \\
\end{eqnarray}
where $B({\bf q}) = \frac{\epsilon_{\bf k} \epsilon^{\prime}_{\bf k} +
{\cal D}_{\bf k}{\cal D}_{\bf k}^{\prime}}{{\cal E}_{\bf k} {\cal E}^{\prime}_{\bf k}}$,
${\cal E}^{\prime}_{\bf k} = {\cal E}_{{\bf k}+{\bf q}}$,
and ${\cal D}^{\prime}_{\bf k} = {\cal D}_{{\bf k}+{\bf q}}$.
Here,
${\cal E}_{\bf k} = 2 \sqrt{\epsilon_{\bf k}^{2} + {\cal D}^{2}_{\bf k}}$ is
the dispersion of the fermionic spinons,\cite{zhou}
$\epsilon_{\bf k} = \lambda \cos(k_x+k_y) - a_3$,
${\cal D}_{\bf k} = \chi (\cos k_x - \cos k_y)$, 
and $f({\cal E}_{\bf k})$ is the Fermi distribution function.
We assume that the values of the order parameter fields $\chi$ and $\lambda$,
which were obtained in Ref.\ \onlinecite{zhou},
do not vary significantly in the temperature range considered here,
and we explicitly verified that our results are insensitive to
any changes in $a_3$ with temperature.


\begin{figure}[h]
\includegraphics[width=8cm,angle=0]{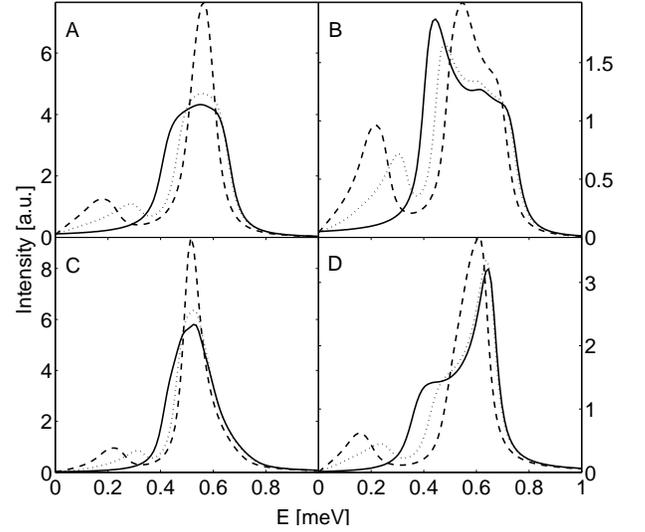}
\caption{Scans of $\chi^{\prime\prime}({\bf q},\omega)$ at 
various temperatures in the bosonic Sp($N$) approach along the scan
directions A, B, C and D as used in Fig.\ 2(a) of Ref.\ \onlinecite{coldea}.
The results at $T = 0$, $0.15 J_2$ and $0.2 J_2$ are shown by the solid,
dotted, and dashed lines, respectively.
}
\label{scanXSpN}
\end{figure}

Our results for $\chi^{\prime\prime}({\bf q},\omega)$ for the two spin-liquid
states are shown in Figs.\ 2-4. 
The spin structure factor ${\cal S}({\bf q},\omega)$
differs from $\chi^{\prime\prime}({\bf q},\omega)$ by an overall thermal factor 
$(1 + n(\omega))$. As a result the low energy spectral weight in 
$\chi^{\prime\prime}({\bf q},\omega)$
is somewhat suppressed compared to
${\cal S}({\bf q},\omega)$, but the overall intensity distribution
looks very similar. 
To make direct contact with experiment, we show 
$\chi^{\prime\prime}({\bf q},\omega)$ for both spin liquid states
in Fig.\ 2,
with ${\bf q}$ oriented along the direction of $J_2$ 
(see Fig.\ 1). Figures 3 and 4 show the same quantity along four scan trajectories 
that were used in the experiment (see Figs.\ 2 and 3 of Ref.\ \onlinecite{coldea}).
Note that the point $k = 2\pi/b$ in
Fig.\ 2a of Ref.\ \onlinecite{coldea} corresponds to $q=\pi$ in Fig.\ 2 here.

\begin{figure}[h]
\includegraphics[width=8cm,angle=0]{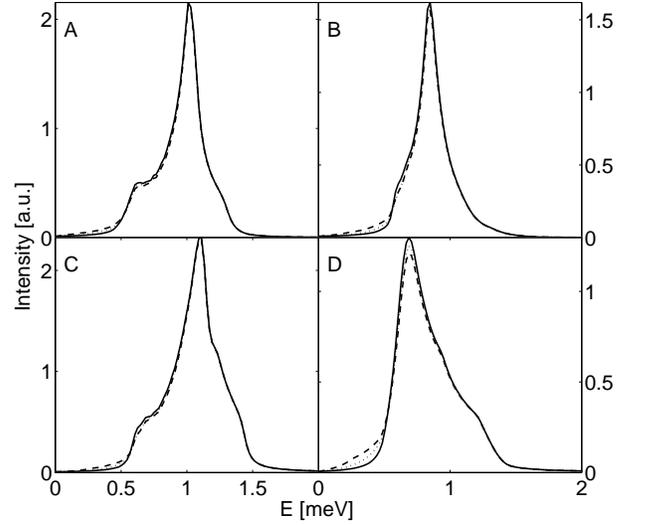}
\caption{Scans of $\chi^{\prime\prime}({\bf q},\omega)$ at 
various temperatures in the fermionic SU(2) approach along the same 
four scan directions as in Fig.\ 3. 
The results at $T = 0$, $0.15 J_2$ and $0.2 J_2$ are shown by the solid, 
dotted, and dashed lines, respectively. 
}
\label{scanXSU2}
\end{figure}

Both spin-liquid states show a continuum of spin excitations,
which is a hallmark of spin-liquids with spin-1/2 spinons. 
Moreover, both theories indicate strong scattering around $(\pi/2,\pi/2)$ in 
agreement with the experiment.  
At zero temperature, the lower edge of the excitation 
spectrum in both theories has minima at $(0,0)$, $(\pi,\pi)$ and 
close to $(\pi/2,\pi/2)$.
The spinon spectrum in the Sp($N$) approach 
has a small gap ($0.03J_2$) at the single incommensurate minimum while 
it has two gapless incommensurate points in the SU(2) theory.  
These slight differences, however, are below the experimental resolution 
(about $0.5 J_2$).\cite{coldea}
The upper boundary of the scattering spectrum in the SU(2) approach
is closer to the experimental results. On the other hand,
the Sp($N$) approach describes some aspects of the
experimentally measured spectra at lower energy (see Fig.\ 2a of
Ref.\ \onlinecite{coldea}) better:
there is much less scattering intensity around $(\pi,\pi)$ in the 
Sp($N$) approach compared to the results of the SU(2) theory, and 
the peak intensity is close to the lower edge as opposed to the upper 
edge in the case of the SU(2) approach.
The details of the spectra described above are expected to change,
however, once  
fluctuations about the mean-field solutions are included. 
In addition, it is difficult to distinguish the two spin-liquid
states from their low energy excitation spectra under current 
experimental resolution. Therefore, it is fair to say that,
at zero temperature, both theories agree 
reasonably well with experiment on a qualitative level.

At finite temperatures, however, one expects that the different spinon
statistics will give rise to substantial differences 
in the spin excitation spectrum.
Indeed, as shown in Figs.\ 2-4, our results at finite temperatures 
indicate that the scattering intensity extends to a broader range in
the phase space of ${\bf q},\omega$, and  
$\chi^{\prime\prime}({\bf q},\omega)$  shows very
different behavior with increasing temperature in the two approaches.
This difference can be most clearly seen by
comparing the temperature evolution of the intensities 
along the four scan directions used in the experiment
(see Fig.\ 2 of Ref.\ \onlinecite{coldea}).  
As shown in Fig.\ 3, the maximum intensity increases in the Sp(N) approach,
and the excitation spectrum becomes more narrowly peaked around the maxima
as temperature increases, while the overall intensity remains roughly 
constant. 
Additionally, a two-peak structure develops, which resembles the lineshape predicted by a 
spin-wave calculation in the LRO phase (see Fig.~3 of Ref.~\onlinecite{coldea}), 
and is presumably due to the vicinity 
of an ordered quantum ground state.
On the other hand, as shown in Fig.\ 4, the evolution of the spin structure
factor with temperature is much less significant in the SU(2) theory,
where only a very slight shift of spectral weight to lower energies is
observed.
Since the different temperature dependence arises from distinct statistics of 
the spinons, these results will not be affected fundamentally by fluctuations 
about the mean-field ground states. 


In summary, we studied the dynamic spin structure factor of different spin-liquid 
states obtained from the bosonic large-$N$ Sp($N$) and the fermionic 
SU(2) mean-field theories of a two-dimensional frustrated spin-1/2
Heisenberg antiferromagnet, as a model for Cs$_2$CuCl$_4$.
We found that at zero temperature the dynamic spin structure factor of 
both spin-liquid states compares favorably with experiment. 
At finite temperatures, fundamentally different behavior arises due to the
different spinon statistics. The signatures we found in the temperature dependence
of the spin structure factor can be used to compare both theories with future 
experiments, in order to determine which theory is best suited to describe
the spin liquid state of Cs$_2$CuCl$_4$. More details will be presented
in a future publication.

We would like to thank J.~B.~Marston and R.~Coldea for stimulating discussions.
This research was supported by the NSERC of Canada, the Sloan
Foundation, and the Canadian Institute for Advanced Research.


\end{document}